\documentclass[epsf,prb,twocolumn,showpacs, superscriptaddress, longbibliography,nofootinbib]{revtex4-1}

\usepackage[pdftex]{graphicx}
\usepackage{dcolumn}
\usepackage{bm}
\usepackage{epsfig}
\usepackage{latexsym}
\usepackage{amsmath}
\usepackage{amsfonts}
\usepackage{amssymb}
\usepackage{color}
\usepackage{array}
\usepackage{framed}
\usepackage[normalem]{ulem}
\usepackage{mathtools}
\usepackage{hyperref}

\newcommand{\Z}{\mathbb{Z}}
\newcommand{\N}{\mathbb{N}}

\begin{document}

\title{Odd Fracton Theories, Proximate Orders, and Parton Constructions}

\author{Michael Pretko}
\affiliation{Department of Physics and Center for Theory of Quantum Matter, University of Colorado, Boulder, CO 80309}

\author{S. A. Parameswaran}
\affiliation{Rudolf Peierls Centre for Theoretical Physics, Clarendon Laboratory, Oxford OX1 3PU, UK}

\author{Michael Hermele}
\affiliation{Department of Physics and Center for Theory of Quantum Matter, University of Colorado, Boulder, CO 80309}

\date{\today}

\begin{abstract}
The Lieb-Schultz-Mattis (LSM) theorem implies that gapped phases of matter must satisfy non-trivial conditions on their low-energy properties when a combination of lattice translation and $U(1)$ symmetry are imposed. We describe a framework to characterize the action of symmetry on fractons and other sub-dimensional fractional excitations, and use this together with the LSM theorem to establish that X-cube fracton order can occur only at integer or half-odd-integer filling.  Using explicit parton constructions, we demonstrate that ``odd'' versions of X-cube fracton order can occur in systems at half-odd-integer filling, generalizing the notion of odd $Z_2$ gauge theory to the fracton setting. At half-odd-integer filling, exiting the X-cube phase by condensing fractional quasiparticles  leads to symmetry-breaking, thereby allowing us to identify a class of conventional ordered phases proximate to phases with fracton order. We leverage a dual description of one of these ordered phases to show that its topological defects naturally have restricted mobility. Condensing pairs of these defects then leads to a fracton phase, whose excitations inherit these mobility restrictions.
\end{abstract}
\maketitle

\normalsize

\section{Introduction}
\label{sec:intro}
A central goal of  condensed matter physics is to understand the low-temperature phase structure of interacting  quantum many-body systems. Historically this study centred on delineating ordered phases based on their distinct symmetry properties. However, over the past three decades an important parallel strand of activity has emerged that focuses instead on more subtle distinctions between quantum-disordered phases that lack the conventional notion of a local order parameter. {Such phases are said to be topologically ordered\cite{wen1989,wen1990,wen1990groundstate,wen2013} when they exhibit fractionalized quasiparticle excitations, which, as exemplified by the fractional quantum Hall effect, can have sharp signatures in experiments.

Despite the rapid progress made in understanding {quantum-disordered phases} in recent years, the field still holds  unexpected surprises that can motivate fundamentally new insights and ideas, or stimulate the development of new theoretical techniques.  A case in point is the identification of a new class of fractionalized phases that exhibit emergent `fracton' quasiparticles with limited mobility.\cite{chamon,haah,fracton1,fracton2,sub,fractonarcmp,review} An individual fracton excitation cannot move by itself, but can move in certain bound states, as dictated by the presence of emergent higher moment conservation laws.\cite{sub,genem,higgs1,higgs2,localization}  Additionally, many systems that give rise to fractons also host ``subdimensional" particles which are only free to move in certain directions. This distinguishes fractons  from  quasiparticle excitations of conventional topological orders, that suffer no  such restrictions, and indicates that the emergent low-energy theory of these models cannot be completely captured by standard topological quantum field theories.  

While fracton phases are now being explored from many different points of view, there is  a need for more clues on how to search for fracton order in experimentally relevant systems.  In part, this is because  most fracton models are designed so as to be exactly solvable or nearly so, and accordingly offer little insight into whether they can indeed emerge as low-energy descriptions of conventional systems of electrons, spins, or bosons. This is to be contrasted with the relatively mature understanding -- especially via so-called parton constructions --  of how such systems can in principle give rise to topologically ordered spin liquid states.\cite{savary2016spinliquids}

We may make progress towards this goal  by {studying the emergence of fracton theories} as long-wavelength effective descriptions of  spin systems with local Hamiltonians. This can also help to identify more conventional ordered phases that are naturally proximate to fracton phases within a specified parameter space.  Experimental searches can then focus on classes of materials whose local energetics favor these proximate orders. Since they are typically equipped with a local order parameter, this is an easier task than directly engineering a fracton phase. Tuning parameters of the system could then enhance the effect of fluctuations to drive a transition into a fracton phase. Such ideas have, for example, identified  proximity to a Mott transition as one physical mechanism that can seed spin liquid behaviour.\cite{motrunich2005variational,lee2005u1} Here, we take the first steps towards developing a similar understanding of fracton physics.

For some types of fracton models, the relevant proximate phases may simply be paramagnets --- defined as gapped quantum-disordered states with no broken symmetries or fractionalized quasiparticles. In such situations, we obtain little experimental guidance.  However, in other cases, we are aided by the Lieb-Schultz-Mattis (LSM) theorem, which guarantees that gapped systems at certain filling fractions are required to exhibit some nontrivial form of ordering.\cite{lsm,hastings,oshikawa} More specifically, the LSM theorem tells us that, if a system at fractional filling is gapped and does not exhibit symmetry-breaking order, it must possess a robust ground state degeneracy, which is one of the signature characteristics of a fractionalized phase.  Systems at fractional filling thereby provide an ideal setting in which to seek experimental realizations of fracton phases and their proximate ordered phases.

In this work, we  derive LSM constraints on a class of gapped fracton phases, and use this to analyse the possible fractonic phases of lattice models with {XY spin rotation and translational symmetry.  (In other words, we consider translation-invariant local bosonic lattice systems with a global $U(1)$ conserved charge.)}  It is worth clarifying what we mean by this at the very outset, in order to place this work in context of recent related results. Several recent works~\cite{HeYouPrem,dubinkin} have also considered LSM-like constraints on fracton models. However, these authors focused on additional LSM-like constraints that can be derived assuming the existence of an additional set of {\it subsystem symmetries}. In contrast, here we consider the `minimal' set of LSM constraints requiring only translational and global $U(1)$ conservation, which are physically relevant to a much larger class of lattice spin systems without requiring fine-tuning (in contrast to subsystem symmetries).   As with more familiar topological phases\cite{arun,glide}, the LSM theorem leads to consistency conditions on what types of fracton phases can be realized at a particular filling. However, as we argue below, the conventional flux-insertion arguments that lead to LSM constraints fail in the fracton case, {motivating a more sophisticated approach that draws on ideas developed in the context of topologically ordered phases ``enriched'' by symmetry, the so-called symmetry enriched topological (SET) phases.}

For concreteness, we focus primarily on variants of the $Z_2$ fracton order realized in the X-cube model\cite{fracton2}, which can be understood as a type of $Z_2$ symmetric tensor gauge theory.\cite{higgs1,higgs2}  The X-cube model contains a variety of emergent quasiparticles of restricted mobility, including immobile fractons, {``planon'' composites of two fractons that move in two-dimensional planes, and ``lineon'' particles restricted to move along certain lines.}  We begin in Section \ref{sec:filling} by demonstrating that, like ordinary $Z_2$ topological order, X-cube fracton order can only be realized at integer or half-integer fillings. {In order to establish this result, we introduce a framework to describe the action of symmetry on the point-like restricted-mobility excitations of fracton phases. We} show that the commonly studied ``even" version of {X-cube order} cannot be consistently realized at half-integer filling.  {Rather, a system at half-integer filling can only exhibit one of several ``odd'' varieties of X-cube fracton order.}  
In Section \ref{sec:odd}, we present a detailed formulation of these odd X-cube theories, which are characterized by a uniform background density of one of the species of emergent quasiparticles.  All nontrivial quasiparticles of these theories carry a fractionalized symmetry quantum number, ensuring that any phase obtained via condensation of quasiparticles will feature some form of symmetry breaking, as required by the LSM theorem.  In Section \ref{sec:parton}, we provide explicit parton constructions for the odd X-cube theories, to demonstrate how they can arise at a microscopic level in models with a spin-$1/2$ site Hilbert space and local interactions. This also demonstrates the inconsistency of the even (odd) fracton theories in models with an odd (even) number of spins-$1/2$ per unit cell.

In the remaining sections, we explore connections between the odd X-cube model and other phases of matter, both through duality arguments and via an analysis of condensation transitions.  In Section \ref{sec:dual}, we construct dualities between  odd X-cube {theories} and various frustrated Ising models.  We first generalize the plaquette Ising duality of the even X-cube model to its odd counterpart.  The corresponding dual of the odd X-cube model is a frustrated version of the plaquette Ising model, with flipped signs on certain plaquette terms in the Hamiltonian.  We also construct a new multi-spin Ising duality for both  even and odd X-cube theories.  In Section \ref{sec:order}, we then consider various ordered phases which can be obtained from the odd X-cube theory via condensation of its emergent quasiparticles.  For example, one odd theory can give rise to a plaquette-ordered phase upon quasiparticle condensation.  We then discuss the converse question of how this plaquette-ordered phase can be driven back into the X-cube phase via condensation of double vortices.  Another odd X-cube theory is proximate to a type of bond order.  This identification of proximate ordered phases provides important hints in the search for fracton phases in experimental settings.  Finally, in Section \ref{sec:conc}, we summarize and discuss some open questions, such as the extension of these concepts to other types of fracton order.

\section{Filling Constraints on Fracton Theories}
\label{sec:filling}

\subsection{Flux Insertion and LSM {Constraints on} Conventional Topological Orders}
\label{sec:conventional-constraints}

The LSM theorem and its various generalizations are rooted in the idea of flux insertion. The basic idea is to consider a finite system with periodic boundary conditions, and examine how the crystal momentum (or other symmetry quantum number) of the ground state of the system changes as a quantum of $U(1)$ flux is threaded through a non-contractible loop of the system. This quantity is insensitive to most microscopic details and depends only on the filling  ($U(1)$ charge per unit cell) and the system size in the direction that encircles the threaded flux. Any putative long-wavelength, low-energy description of the system must be consistent with this `symmetry inflow' (in the thermodynamic limit) and is therefore highly constrained.

 As an example, consider a system of $N$ charge-1 particles (either bosons or fermions) on an $L_x$-by-$L_y$ square lattice, wrapped into a cylinder along the $L_x$ direction.  We then adiabatically~\footnote{Technically speaking this should be a quasi-adiabatic insertion, but we will not dwell on this subtlety here; see Refs.~\onlinecite{hastingsSufficient,HastingsQA,hastings} for details.} insert a magnetic flux of $2\pi$ through the hole of the cylinder, as depicted in Fig.~\ref{fig:cyl}, with the flux $\Phi(t)$ slowly ramping up from $0$ to $2\pi$.  By Faraday's law, the electric field along the $x$-direction of the lattice is given by:
\begin{equation}
E_x = \frac{1}{L_x}\frac{d\Phi}{dt}
\end{equation}
and the total momentum imparted to the system is given by:
\begin{equation}
\Delta P_x = N\int dt \frac{1}{L_x}\frac{d\Phi}{dt} = -\frac{2\pi}{L_x}N.
\end{equation}
It is now useful to rewrite this Eq.~using the filling fraction, $\nu = N/L_xL_y$, in terms of which we have:
\begin{equation}
\Delta P_x = 2\pi \nu L_y.
\end{equation}
(Note that we are free to choose $N$ to be a multiple of $L_x$, so that the total charge remains an integer, as it must). A similar result holds for $\Delta P_y$ if the system is wrapped into a cylinder along the $L_y$ direction. 
\begin{figure}[t!]
 \centering
 \includegraphics[scale=0.44]{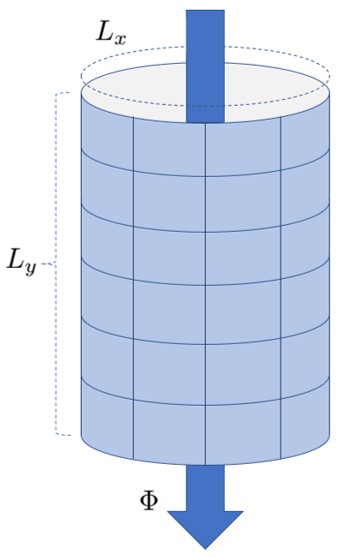}
 \caption{We consider a system in a cylindrical geometry, in which a flux $\Phi$ is threaded through the hole of the cylinder.  By studying the resulting change in momentum as a function of filling, we obtain restrictions on the corresponding phase of matter.  Fig.~adapted from Ref.~\onlinecite{arun}.}
 \label{fig:cyl}
\end{figure}

If we assume that the system has an energy gap and a unique ground state, then such a flux-insertion process should leave the ground state invariant.  This is consistent only if $\Delta P_x$ is a multiple of $2\pi$ (since crystal momentum is defined mod $2\pi$). In other words, a unique ground state is consistent only if $\nu L_y$ is an integer.  When the filling $\nu$ is itself an integer, this condition is always satisfied.  For fractional fillings, however, a unique ground state is only allowed at certain values of $L_y$.  This indicates that, in the thermodynamic limit, the system must either have degenerate ground states, or the assumption of an energy gap must be false.  In the former case, this degeneracy can arise either from spontaneous symmetry breaking or (in $d>1$) from topological order.  If a system at fractional filling is known to be both gapped and symmetric, then the only remaining possibility is a topologically ordered phase.

In addition to dictating when a system must exhibit topological order, this flux-insertion argument can also constrain the types of topological order which can occur at particular fillings.\cite{arun,glide}  For instance, consider a system at filling $\nu = p / q$, with $p$ and $q$ relatively prime integers.  Choosing $L_x$ divisible by $q$ (so that the total charge is an integer), and $L_y$ relatively prime to $q$, the flux insertion argument implies the existence of $q$ degenerate ground states, with crystal momenta differing by multiples of $\Delta P_x = 2\pi \nu L_y$.  In fact, because the adiabatic flux insertion is unitary, it follows that the total ground state degeneracy is divisible by $q$, because the subspace of ground states at a fixed crystal momentum $P_x$ is unitarily mapped to a corresponding subspace with crystal momentum $P_x + \Delta P_x$.  This constrains which topological orders  can occur in principle at filling $\nu = p / q$.

There are also other types of constraints originating from the LSM theorem.  To illustrate this,} recall that a 2D $Z_2$ gauge theory has three types of gapped quasiparticle excitations: bosonic electric charges ($e$), magnetic fluxes ($m$), and their fermionic bound state ($\epsilon\equiv e\times m$). {Suppose we consider a system at half-filling, where a non-trivial LSM constraint holds,} and choose the $e$ particles to carry the fractionalized $U(1)$ charge of the theory, and the $m$ particles to be charge neutral.  If the $m$ particles do not carry any other nontrivial quantum numbers, then condensing them would lead to a trivial symmetric gapped phase, which is not possible at half-filling.  We therefore conclude that the $m$ particles must carry some fractional quantum number.  {While by assumption they do not carry the $U(1)$ charge of the theory, it is possible for excitations to carry fractional crystal momentum,\cite{wen02,EssinHermele} and this is the only remaining possibility.

The condition that the $m$ particles carry fractional crystal momentum has consequences for the form of the effective pure $Z_2$ gauge theory that describes the physics below the gap to electrically charged $e$ excitations.  Specifically, this gauge theory is of the ``odd'' variety\cite{MoessnerSondhiFradkin}, meaning there is a background $Z_2$ charge on every site of the lattice.  We illustrate this on the square lattice, where the $Z_2$ gauge degrees of freedom are spin-$1/2$ spins on nearest-neighbor links labeled by $\ell$, subject to the Gauss' law constraint
\begin{equation}
A_v \equiv \prod_{\ell\in +_v}\sigma^z_\ell = -1 \text{.}  \label{eqn:oddgauss}
\end{equation}
Here, $v$ is a vertex of the square lattice, and the product of $\sigma^z_\ell$ Pauli operators is over the four links adjacent to $v$.  The ``$-1$'' on the right-hand side of Eq.~(\ref{eqn:oddgauss}) indicates the presence of a background ``electric'' $Z_2$ gauge charge on the vertex $v$.  An exactly solvable Hamiltonian for the deconfined phase of the odd $Z_2$ gauge theory is
\begin{equation}
H_{\text{odd-gauge}} = - \sum_p B_p  \text{,}
\end{equation}
where the  sum is over plaquettes $p$ and $B_p=\prod_{\ell\in\square_p} \sigma_\ell^x$.

The odd $Z_2$ gauge theory with Hamiltonian $H_{\text{gauge}}$ is closely related to the ``odd toric code'' Hamiltonian
\begin{equation}
H_{\text{TC}, o} = \sum_v A_v - \sum_p B_p \text{.}
\end{equation}
Here again spin-$1/2$ spins reside on each link of the square lattice, with $A_v$ and $B_p$ defined as above.  Unlike the odd $Z_2$ gauge theory, no local constraint is imposed on the Hilbert space.  However, if we project to the subspace where $A_v = -1$, which minimizes the first term in the Hamiltonian, then the odd toric code becomes equivalent to the odd $Z_2$ gauge theory.

The $A_v = -1$ background charge implies that an $m$ particle picks up a minus sign upon going around a single vertex.  In other words, if we define $T^{(m)}_x$ and $T^{(m)}_y$ as the operators which translate a single localized $m$ particle by one lattice constant in the $x$ and $y$ directions, respectively, we have:
\begin{equation}
T^{(m)}_xT^{(m)}_y[T^{(m)}_x]^{-1}[T^{(m)}_y]^{-1} = -1.
\end{equation}
The fact that the $m$ particles transform projectively under translations is what it means for the crystal momentum to be fractional.  Upon condensation of the $m$ particles, this fractionalized momentum will cause the system to develop some type of spatial order.  In this way, the odd nature of the $Z_2$ gauge theory description rescues the system from violating the LSM theorem.

\subsection{LSM {Constraints on} Fracton Orders\label{sec:lsmfracton}}
We have reviewed how {the LSM theorem places} a set of restrictions on {topologically ordered} phases. We now discuss analogous constraints for fracton orders, focusing on the so-called X-cube fracton order, realized in the exactly solvable X-cube model.\cite{fracton2}  This is a model of spin-$1/2$ spins on the links of the simple cubic lattice, with the Hamiltonian a sum of commuting terms given by
\begin{equation}
H_{\text{X},e}  = -\sum_{v\mu } A^v_{\mu}  - \sum_c B_c.
\label{xcube}
\end{equation}
(The notation differs slightly from that of Ref.~\onlinecite{fracton2}.)  The last term represents a sum over all cubes of the lattice, where  $B_c$  is defined as a product of the twelve $\sigma^x_\ell$'s on the boundary of a cube:
\begin{equation}
B_c = \prod_{\ell\in\partial c} \sigma^x_\ell.
\end{equation}
In the first term, the sum on $v$ runs over all vertices, while the sum on $\mu$ runs over the three Cartesian coordinate directions $x$, $y$ and $z$.  Each $A^v_{\mu}$ term involves only the four coplanar links lying in the plane orthogonal to $\mu$, out of the six total links touching $v$.  For example,  
$A^v_{z}$  is a product over the four links in the $xy$-plane touching $v$:
\begin{equation}
A^v_{z} = \prod_{\ell\in+_{xy, v}} \sigma^z_\ell.
\end{equation}
Note that the three $A^v_\mu$ operators on a given vertex obey the important relation:
\begin{equation}
A^v_{x}A^v_{y} = A^v_{z}.
\label{relation}
\end{equation}
The terms of the Hamiltonian are illustrated in Fig.~\ref{fig:xcube}.

{Ground states of the X-cube model are eigenstates of all the $A^v_\mu$ and $B_c$ operators, with eigenvalue $+1$. There are two types of point-like quasiparticles, which are elementary in the sense that arbitrary excitations can be constructed as composites of the elementary excitations.  First, cubes with $B_c = -1$ are fractons, which are individually immobile.  However pairs of fractons separated along one of the coordinate axes are mobile in a plane normal to the axis, and are thus examples of ``planon'' excitations.  This can be seen by studying the subsystem conservation laws of Eq.~\eqref{xcube}, which imply that the fracton number is conserved modulo two on planes normal to the coordinate directions.}
 Second, a vertex with  
 $A^v_{z} = -1$, $A^v_{y}=-1$, and $A^v_{x} = 1$
 is a lineon with mobility only in the $x$ direction, and similarly for the other directions.  (Note that we cannot have only a single $A^v_{\mu}$ operator at $v$ with negative eigenvalue, due to the restriction of Eq.~\eqref{relation}.)
 \begin{figure}[t!]
 \centering
 \includegraphics[scale=0.4]{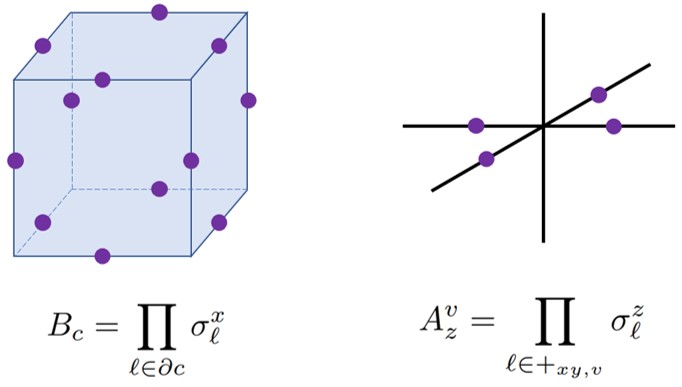}
 \caption{The Hamiltonian of the X-cube model contains two types of terms: (left panel) a product $B_c$ of $\sigma^x_\ell$ over the twelve links in the boundary of a cube $c$, and (right panel) products $A^v_{\mu}$ over four coplanar spins touching each vertex $v$, shown for $\mu = z$.}
 \label{fig:xcube}
\end{figure}

In this section, we argue that the LSM theorem implies that X-cube fracton order is only consistent at integer and half-integer fillings.  (By ``half-integer'' here and elsewhere we mean more precisely ``half-odd-integer.'')  While flux insertion would seem to provide a route to such a constraint,  a subtle but serious obstruction to  such an argument arises.  In a three-dimensional $L_x \times L_y \times L_z$ system at filling $\nu$, adiabatically inserting flux in the $x$-direction results in a crystal momentum change of $\Delta P_x = 2 \pi \nu L_y L_z$.  If $\nu = 1/2$, we thus need to choose the product $L_y L_z$ to be odd in order to obtain a non-trivial constraint.  However, this relies on the assumption that there exists a gapped ground state for system sizes approaching the thermodynamic limit with $L_y L_z$ odd.  This assumption turns out to be problematic for X-cube fracton order, as we demonstrate in Appendix~\ref{app:LSMfail}.  There we describe a modified X-cube model that, by construction, can arise in a $\nu = 1/2$ system, but which does not admit a gapped ground state when any two of $L_x$, $L_y$ or $L_z$ are odd.  Indeed, it is a familiar fact that 2D odd $Z_2$ gauge theory suffers a similar problem in an odd-by-odd system.  Given the close relationship between the X-cube model and a system of decoupled layered 2D $Z_2$ gauge theories\cite{layer1,layer2}, the result of Appendix~\ref{app:LSMfail} can be anticipated.

We instead take an alternative approach to deriving LSM constraints, that rests on the idea that one can combine topological phases by stacking them and condensing bound states of fractionalized excitations to access new phases. The basic idea in the current setting is to consider a system with X-cube order at filling $\nu$, and stack this system with itself to obtain a new system at filling $2 \nu$.  We will argue that in this new system we can always condense particle-like excitations to obtain a trivial gapped phase, which implies that $2 \nu$ is an integer.  Therefore X-cube order can only occur at half-integer filling $\nu$. We note that similar arguments  also apply for more conventional topologically ordered theories, and present an alternative to the usual flux-insertion arguments.\cite{arun}

To proceed we consider the symmetry group $G = \Z^3 \times {\rm U}(1)$, where $\Z^3$ is the group of lattice translations.  We will need to characterize the \emph{symmetry-enriched fracton} (SEF) order of systems with X-cube order and  the $G$ symmetry.  The perspective will be to first characterize the X-cube order, and then describe how this order is enriched by the $G$ symmetry, separating out the description of the fracton order from that of its symmetry enrichment.  The same perspective is often employed in more conventional symmetry-enriched topological (SET) phases, which are topologically ordered phases (\emph{i.e.} non-invertible topological phases) in the presence of symmetry.\cite{WenPSG,EssinHermele,MesarosRan,Gcrossed}  Based on the now-extensive understanding of SET phases, we assume that SEF phases can be completely characterized by {describing the fractional excitations and the action of symmetry on fractional excitations.}  In fact, such a characterization -- unless one also considers extrinsic defects, \emph{i.e.} symmetry fluxes -- is only expected to give a complete description of SET phases up to stacking with a symmetry-protected topological (SPT) phase, or other invertible topological phase, an issue that also arises for SEF phases.  However, we can safely ignore this issue as it does not affect LSM constraints; invertible topological phases have a unique ground state on the torus, and can thus only occur at integer filling in systems with $\Z^3 \times {\rm U}(1)$ symmetry.

We restrict attention to Abelian fracton orders with only point-like fractional excitations, which includes the X-cube order of interest.  Particle types are labeled by elements of an Abelian group ${\cal A}$, where the addition operation in ${\cal A}$ corresponds to fusion of particles.  The same description applies to Abelian topological orders in two dimensions, where ${\cal A}$ is a finite group.  However, as emphasized in Ref.~\onlinecite{fusion}, this group is not finitely generated in fracton orders; in the X-cube order, ${\cal A}$ is {not finitely generated, and can be constructed as a quotient of a countably infinite} direct sum of $\Z_2$ summands.

A key difference between SEF and SET phases is the presence of excitations with restricted mobility in the former.  In order to have a workable characterization of fracton orders in terms of their excitations, the restricted mobility should be incorporated somehow.  This can be done in a precise way using lattice translation symmetry, which acts on ${\cal A}$ by permuting particle types.\cite{haah13commuting,haah16algebraic,fusion}  Mathematically, this action is described by a homormophism $\rho : \Z^3 \to \operatorname{Aut}({\cal A})$.  Given the crucial importance of restricted-mobility excitations, and the lack of a precise way of describing them apart from lattice translation symmetry, we take the following point of view:  \emph{the action on ${\cal A}$ by translation symmetry, as specified by $\rho$, is part of the specification of a fracton order itself.}  This point of view contrasts with that taken in describing SET phases, where a clean separation between the description of a topological order, and the action of symmetry on its excitations, is possible.  We then take the X-cube order to be defined by the example of the X-cube model, with the action $\rho$ given by the full translation symmetry of the cubic lattice.  Mathematically, $\rho$ makes the Abelian group ${\cal A}$ into a module over $\Z[\Z^3]$, which is described in detail for the X-cube order in Ref.~\onlinecite{fusion}.  To summarize some key points, ${\cal A} = {\cal A}_f \oplus {\cal A}_\ell$, where ${\cal A}_f$ consists of all particle types obtained as composites of fractons, and ${\cal A}_\ell$ is similar but for the lineon excitations.  ${\cal A}_f$ has a single generator, denoted $f$, which is the particle type of a fracton at some arbitrary fixed position (other fractons are obtained from $f$ by acting with translation).  Similarly, ${\cal A}_\ell$ has two generators, which can be chosen as $\ell_x$ and $\ell_y$, corresponding to lineons moving in the $x$- and $y$-directions, respectively, at some arbitrarily chosen locations.

Before turning to the description of the SEF X-cube order, we first review the corresponding description in an Abelian SET phase with unitary symmetry $G$, focusing on two dimensions where all excitations are point-like.  See Appendix~\ref{app:math} for additional mathematical detail that is covered lightly in the main text to simplify the presentation.  The action of symmetry on fractional excitations is specified by two pieces of data.  First, the symmetry may permute particle types, as specified by a homomorphism $\rho : G \to \operatorname{Aut}({\cal A})$.  This makes ${\cal A}$ into a $\Z[G]$-module, or a $G$-module for short.  Second, the symmetry fractionalization is specified by an element $[\omega]$ of the second cohomology group $H^2 (G, {\cal A}^*)$.  Here,  ${\cal A}^* = \operatorname{Hom}({\cal A},{\rm U}(1))$ is the group of homomorphisms from ${\cal A}$ to ${\rm U}(1)$, sometimes referred to as the Pontryagin dual of ${\cal A}$.  ${\cal A}^*$ inherits a $G$-module structure from that of ${\cal A}$, and this structure enters into the definition of the cohomology group; see Appendix~\ref{app:math} for details.  When ${\cal A}$ is finite, we have ${\cal A} \cong {\cal A}^*$ (see Appendix~\ref{app:math}), so often a distinction is not made between ${\cal A}$ and ${\cal A}^*$.

A point not often emphasized, but one that will be important for applications to fracton orders, is that the physical interpretation of $[\omega]$ differs depending on whether ${\cal A}$ or ${\cal A}^*$ is chosen as the coefficient group.  In either case, we let $\omega(g_1, g_2)$ be a specific 2-cocycle representing the class $[\omega]$.  With ${\cal A}^*$ coefficients, $\omega(g_1, g_2)$ is a homomorphism from ${\cal A}$ to ${\rm U}(1)$, whose values can be written $\omega_a (g_1 ,g_2) \in {\rm U}(1)$, for $a \in {\cal A}$.  In the simple case where $\rho$ is trivial, \emph{i.e.} when symmetry does not permute particle types, $\omega_a(g_1, g_2)$ has a simple physical interpretation having to do with symmetry localization.\cite{EssinHermele,MesarosRan} Suppose a state $| \Psi \rangle$ contains anyon excitations of types $a_1, \dots, a_n$, well-separated from one another in space, and that the system is locally in the ground state away from these excitations.  Then if $g \in G$ is represented by the unitary $U(g)$, we have
\begin{equation}
U(g) | \Psi \rangle = U_{a_1}(g) \cdots U_{a_n}(g) | \Psi \rangle \text{,}
\end{equation}
where $U_{a_i}(g)$ has support in a bounded region surrounding the excitation $a_i$.  These operators obey the algebraic relations
\begin{equation}
U_{a_i} (g_1) U_{a_i} (g_2) = \omega_{a_i} (g_1, g_2) U_{a_i} (g_1 g_2) \text{,}
\end{equation}
and we see that $\omega_a(g_1, g_2)$ enters as the so-called factor system of a projective representation describing the symmetry action on $a$.  A similar interpretation holds, though is more subtle, even when $\rho$ is non-trivial;\cite{Gcrossed}  the details will not be needed for our purposes.

If instead we choose the coefficient group to be ${\cal A}$, the interpretation of $\omega(g_1, g_2)$ is in terms of fusion of symmetry fluxes.\cite{chen14anomalous}  We denote a point-like $g$-flux by $\Omega_g$.  This is a point-like extrinsic defect, and more specifically is a gauge flux of a non-dynamical $G$ background gauge field.    We have the projective fusion rule
\begin{equation}
\Omega_{g_1} \Omega_{g_2} = \omega(g_1, g_2) \Omega_{g_1 g_2} \text{.}
\end{equation}
The interpretation is that fluxes fuse according to the group multiplication in $G$, as they must, but only up to fusion with an anyon given by $\omega(g_1, g_2) \in {\cal A}$.  While this interpretation clearly differs from that associated with ${\cal A}^*$ coefficients, they are of course related, with $\omega_a(g_1, g_2) \in {\rm U}(1)$ being the mutual statistical phase for a process where $a$ is braided counterclockwise around $\omega(g_1, g_2) \in {\cal A}$.  

Now we extend this discussion to X-cube SEF order.  The first point is that we need to choose the coefficient group to be ${\cal A}^*$ rather than ${\cal A}$.  Unlike in two-dimensional SET phases, where this choice is a matter of different physical interpretation, here it is more important, because ${\cal A}$ and ${\cal A}^*$ are not in general isomorphic when these groups are not finitely generated.  (See Appendix~\ref{app:math} for more detail.)  We choose ${\cal A}^*$, because the physical interpretation of symmetry fractionalization in terms of symmetry localization is still valid for a fracton order with point-like excitations.  On the other hand, the interpretation in terms of projective fusion of symmetry fluxes is no longer valid, because in three-dimensions the fluxes are line objects and are not point-like.

Next, we specialize to the relevant symmetry group $G = \Z^3 \times {\rm U}(1)$.  As a continuous group, the ${\rm U}(1)$ symmetry cannot permute particle types, and therefore $\rho : G \to \operatorname{Aut}({\cal A})$ is specified entirely by the action of translation symmetry $\rho : \Z^3 \to \operatorname{Aut}({\cal A})$, which is specified as part of the fracton order.  Therefore the SEF data, as distinguished from the data characterizing the fracton order, is given completely by an element $[\omega] \in H^2 (G, {\cal A}^* )$. An important point will be that $\omega_a(g_1, g_2)$ can only take values $\pm 1$, which follows from the fact that any particle type in the X-cube order fuses with itself to the trivial type $0 \in {\cal A}$.  More formally, $a + a = 0$ for any $a \in {\cal A}$, and $\omega_{a_1}(g_1, g_2) \omega_{a_2}(g_1, g_2) = \omega_{a_1 + a_2}(g_1, g_2)$, which implies $[\omega_a (g_1, g_2) ]^2 = 1$.

Now we suppose we have a system with SEF X-cube order at filling $\nu$, and stack this system with itself.  The filling of the new system is of course $2 \nu$, and it has two copies of X-cube order.  The group of particle types ${\cal A}$ is a direct sum ${\cal A} = {\cal A}_1 \oplus {\cal A}_2$, with the summands corresponding to the two copies of X-cube order.  We denote the generators of the fracton sectors of ${\cal A}_1$ and ${\cal A}_2$ by $f_1$ and $f_2$, and similarly denote by $\ell_{1x}, \ell_{1y}, \ell_{2x}, \ell_{2y}$ the generators of the lineon sectors.

We consider condensing composite excitations $f_1 + f_2$.  These are bound states of a fracton in ``layer 1'' with another in ``layer 2,'' at the same spatial positions.  It is easy to see that it is possible to condense such excitations -- this can be done starting with two copies of the standard X-cube model, adding a term $-h \sum_{\ell} \sigma^{z}_{\ell 1} \sigma^z_{\ell 2}$ to the Hamiltonian, where the second index on the Pauli operators is a layer index. We make $h$ sufficiently large, so that $f_1 + f_2$ bound states proliferate and condense.  Upon such condensation, one obtains a new system with a single copy of X-cube order, with fracton sector generated by $f = f_1 \cong f_2$, \emph{i.e.} in the presence of the $f_1 + f_2$ condensate, the distinction between $f_1$ and $f_2$ fractons goes away.  In the lineon sector, the generators $\ell_{1x}, \ell_{1y}, \ell_{2x}, \ell_{2y}$ are all confined by the condensate, but $\ell_x = \ell_{1x} + \ell_{2x}$ and $\ell_y = \ell_{2x} + \ell_{2y}$ are new lineon generators that do not feel the condensate and remain deconfined.

The question then is whether $f_1 + f_2$ can be condensed without breaking symmetry.  Clearly this can be done in the standard X-cube model, via the $-h \sum_{\ell} \sigma^{z}_{\ell 1} \sigma^z_{\ell 2}$ term above.  The key point is that whether or not $f_1 + f_2$ can be condensed without breaking symmetry should depend only on the symmetry fractionalization data of $f_1 + f_2$ excitations, \emph{i.e.} on $\omega_{f_1 + f_2}(g_1, g_2)$. 
But $\omega_{f_1 + f_2}(g_1, g_2) = \omega_{f_1}(g_1, g_2) \omega_{f_2}(g_1, g_2) = [\omega_{f_1}(g_1, g_2)]^2 = 1$, because the two layers have the same SEF X-cube order.  Therefore $f_1 + f_2$ always carries the same (trivial) SEF data as in the case of stacking two standard X-cube models, and is always condensible without breaking symmetry.

We do not yet have a gapped trivial phase, but we can obtain one by condensing the new lineons $\ell_x$ and $\ell_y$. These arise as bound states of lineons in the two layers, carrying identical SEF data, so by the same argument as above, $\ell_x$ and $\ell_y$ carry trivial SEF data.  To see that this is enough to be able to condense the lineons without breaking symmetry, one need only consider the standard X-cube model, now adding the term $-h \sum_{\ell} \sigma^x_{\ell}$, where lineons proliferate and condense once $h$ is sufficiently large.  The remaining fracton excitation $f$ is confined in the presence of the lineon composite, and a gapped trivial phase results, implying that $2 \nu$ is an integer and thus $\nu$ is a half-integer.

A very important point is that if a given SEF X-cube order is possible for integer filling, it is impossible for half-integer filling, and vice versa.  This means that for every SEF X-cube order, there are three mutually exclusive possibilities:  (1) the SEF order occurs for integer filling, (2) the SEF order occurs for half-integer filling, or (3) the SEF order cannot occur in a strictly three-dimensional system (but may potentially occur at the boundary of a four-dimensional system).  We are only concerned with the first two possibilities in this work; the latter possibility may be interesting to explore in the future.  To obtain this conclusion, suppose that the same SEF X-cube order occurs in two different systems, one at integer filling $\nu_1$, and the other at half-integer filling $\nu_2$; we will obtain a contradiction.  Stacking these two systems produces a new system at filling $\nu = \nu_1 + \nu_2$, which is a half-integer.  We can run the same argument as above to condense excitations and obtain a gapped trivial phase, which contradicts the LSM theorem.

A good question at this point is how much the above conclusions depend on our point of view that the X-cube order includes the action $\rho$ of translation symmetry on fractional excitations.  This question is challenging to address, because it is not clear how much freedom one has to change $\rho$ while still maintaining a fracton order that can sensibly be viewed as X-cube order.  However, one simple observation is that we can start from the standard X-cube model, and then lower the translation symmetry to an arbitrary subgroup that is still isomorphic to $\Z^3$.  This corresponds to enlarging the crystalline primitive cell, and results in a different choice of $\rho$.  All the arguments then run almost exactly as above, with no difference in the conclusions -- the key fact is that we still have $[\omega_a(g_1, g_2)]^2 = 1$ for any particle type $a \in {\cal A}$ of such a modified X-cube fracton order.

\section{Odd X-Cube Models}
\label{sec:odd}

The conventional ($i.e.$ ``even") X-cube model is canonically formulated in terms of spin-$1/2$'s on each link of a cubic lattice.\cite{fracton2}  The Hamiltonian for this model was provided in Eq. (\ref{xcube}), and a summary of the quasiparticle types was provided in Sec.~\ref{sec:lsmfracton}.

We now seek to formulate ``odd" versions of the X-cube model in which quasiparticles carry fractionalized crystal momentum in some sense.  {These models are related to odd generalized gauge theories, which we will show in the following section can arise as low-energy effective descriptions in systems at half-filling.  The relationship is analogous to that between the odd toric code and odd $Z_2$ gauge theory discussed in Sec.~\ref{sec:conventional-constraints}.}

  {As for the odd toric code model}, these theories can be constructed by flipping the signs of certain terms of the Hamiltonian.  For example, we can flip the sign of two of the $A$ terms of the Hamiltonian to yield:
\begin{equation}
H_{\text{X},oyz} = \sum_{v} (A^v_{y} + A^v_{z}-A^v_{x}) - \sum_c B_c.
\label{odd}
\end{equation}
In this case, $A^v_{y}$ and $A^v_{z}$ prefer to be in their $-1$ state, while it is still favorable for $A^v_{x}$ to be in the $+1$ state. (We use the {subscript} {`$oyz$'} to denote it is an odd theory where the $y, z$ vertex terms are flipped; the generalization of the notation is obvious.)  This type of state corresponds to having a background density of $x$-directed lineons, with one lineon on each vertex.  Similarly, we could have considered flipping other combinations of two $A$ terms in order to obtain a background density of $y$- or $z$-directed lineons.  We thereby obtain an intuitive understanding of three different types of odd X-cube models, corresponding to uniform background densities of the the three types of lineons.  Since any two different types of lineons can fuse to form the third type, there are no further possible configurations of lineon charge backgrounds.  Note that we also could have considered a model in which only one of the $A$ terms had a flipped sign.  In this case, however, there is no configuration of spins which allows all $A$ and $B$ terms to have their energetically preferred value, and the resulting model is frustrated.  We will not consider such models further in this paper.

For concreteness, we now focus on the odd X-cube model of Eq.~\eqref{odd}.  {We emphasize that this model is only consistent with LSM constraints at integer filling.  More precisely, if we assume a $U(1)$ global symmetry that acts trivially on the spin-$1/2$ degrees of freedom of the model, and if we assume that translations act without any internal rotation in spin space (\emph{i.e.} a translation $t$ sending $\ell \mapsto t \ell$ acts on Pauli operators by $\sigma^\mu_\ell \mapsto \sigma^\mu_{t \ell}$), then it is possible to condense lineon excitations to obtain a trivial gapped phase.  This is so because the lineons carry the same SEF data as in the ordinary even X-cube model, because the ground state has no background of fractons.  Another even simpler way to reach the same conclusion is to observe that in a large Zeeman field, the model enters a gapped trivial phase.}

{The situation changes upon projection to the subspace where the $A^v_\mu$ terms in the Hamiltonian are minimized. This results in a generalized gauge theory with Gauss law constraint $A^v_y = A^v_z = -1$ (this implies $A^v_x = 1$).  This theory thus features a uniform background of $x$-directed lineons.}
This background density leads to important consequences for the fracton sector of the theory, manifesting most clearly in the behavior of {planon composites of two fractons.  Planons moving normal to} the $y$- or $z$-directions have mutual semionic statistics  with $x$-directed lineons within the same plane of motion.  (Note that this sense of mutual statistics is well-defined, even in this three-dimensional system, due to the restricted two-dimensional motion of {the planons}.)  {Planons moving normal to the $y$- or $z$-direction} will then pick up a phase factor of $-1$ upon going around any vertex of the lattice.  As in our previous discussion of the toric code, this phase factor indicates that these {planons} carry a fractional crystal momentum.  As such, condensing these {planons} will naturally lead to a spatially ordered phase, the precise form of which we consider later.  
These considerations suggest that this odd X-cube gauge theory can arise in a system at half-integer filling, which we show by an explicit parton construction in the following section.  The theory avoids running afoul of the LSM theorem, since destruction of the fracton order, {\emph{e.g.} driven by condensation of planons}, will coincide with the development of spatial symmetry breaking. While particles in the fracton sector carry fractional crystal momentum, we should expect that lineon excitations, once included in the theory, carry fractional charge under the $U(1)$ global symmetry, an expectation born out by the parton construction in Section~\ref{sec:parton}.
  
Finally, we consider  a separate type of odd X-cube model in which we flip the sign on the $B$ term of the X-cube Hamiltonian, while keeping the $A$ terms the same:
\begin{equation}
H_{\text{X},oc}  = -\sum_{v,\mu } A^v_{\mu} + \sum_c B_c
\label{eq:odd2}
\end{equation}
where we use the subscript `$oc$' to denote the fact that it is an odd theory obtained by flipping the cube term. (We will occasionally refer to this as the `odd-cube' model.)  Again, we project to the subspace $B_c = -1$, which minimizes the contribution to the energy of the cube term, corresponds to a uniform background of fractons, and results in a distinct type of odd generalized gauge theory.  In this case, we expect that the lineon excitations should carry some type of fractionalized momentum quantum number, which we can see by studying ``dipoles'' of lineons.  
For example, consider a bound state of two $x$-directed lineons separated {by a single lattice constant} along the $z$-direction, which is a {planon}  moving only within the $xy$-plane.  This lineon dipole has mutual $\pi$ statistics with a fracton within its plane of motion, meaning that the dipole acquires a phase factor of $-1$ upon going around any cube of the lattice.  As in previous examples, this implies that the lineon dipole carries fractional crystal momentum.  Any topologically trivial gapped phase obtained by condensing lineons will therefore break spatial symmetries, {which is consistent with this generalized gauge theory arising at half-integer fillings.}

To summarize the results of this section, a consistent way for X-cube fracton order to arise at half-integer filling is for particles either in the fracton or the lineon sector to carry fractional crystal momentum.  This occurs in the two different classes of odd X-cube theories that we formulated here, which correspond to uniform background densities of either lineons or fractons.  In the former case, there are three different odd theories corresponding to the three different orientations of the lineon background, while particles in the fracton sector -- specifically, planon composites of two fractons -- carry fractional crystal momentum.  In the latter case, a uniform background of fractons induces non-trivial momentum quantum numbers in the lineon sector, which we exposed by considering dipoles of lineons that move within two-dimensional planes.

\section{Parton Constructions}
\label{sec:parton}

{Thus far, we have not yet linked the odd X-cube models and corresponding generalized gauge theories to physical systems at half-integer filling.  We now do this by extending the generalized gauge theories constructed in the preceding section to include dynamical matter degrees of freedom.  The resulting parton theories have a deconfined phase with X-cube fracton order, and a confining limit where the Hilbert space reduces to that of a spin-$1/2$ spin model with a global XY (or Heisenberg) spin symmetry, and an odd number of spin-$1/2$ moments per crystalline unit cell.  The existence of such a confining limit shows that the generalized gauge theories can emerge as low-energy effective description of the corresponding spin model.  In the deconfined phase, below the gap to excitations carrying the fractional $U(1)$ charge, the parton theories reduce to the pure generalized gauge theories introduced previously.} We also go further and perturb around the confining limit to extract simple effective models that contain terms that we might expect to stabilize a fracton phase of the corresponding local spin model.

{We first consider the case of the generalized pure gauge theory with $B_c = -1$, constructed in Sec.~\ref{sec:odd} from the model  $H_{\text{X}, oc}$ defined in (\ref{eq:odd2}).  The parton generalized gauge theory again has spin-$1/2$ spins on links of the cubic lattice, and also includes fermionic partons $f_{c \alpha}$ placed on the center of each cube $c$.  The choice of fermions as opposed to bosons is not important and is made purely for technical convenience.  The fermions carry spin-$1/2$, with $\alpha = \uparrow,\downarrow$ the spin index, and transform under a global XY (\emph{i.e.} $U(1)$) spin-rotation symmetry (rotations about the $z$-axis in spin space).  Therefore the $f_{c \alpha}$ fermions are fractionally charged under the $U(1)$ global symmetry, with $\alpha = \uparrow$ ($\alpha = \downarrow$) fermions carring charge $+1/2$ ($-1/2$). None of the conclusions are affected if we enlarge the XY spin-rotation symmetry to the full ${\rm SU}(2)$ Heisenberg symmetry.  The gauge constraint is modified to be $\hat{G}_c = 1$, with
\begin{equation}
\hat{G}_c =  (-1)^{\hat{n}_c + n^0_c} B_c \text{,}
\end{equation}
with $\hat{n}_c = \sum_\alpha f^\dagger_{c \alpha} f^{\vphantom\dagger}_{c \alpha}$ the fermion number at $c$ and $n^0_c$ a static background gauge charge, which we take to be $n_c^0 = 1$.}

We consider the Hamiltonian
\begin{equation}\label{eq:Hgaugec}
H_{\text{gauge},c} = -\sum_{v, \mu } A^v_{\mu} + U \sum_{c, \alpha} f^\dagger_{c\alpha} f^{\phantom\dagger}_{c\alpha} - h\sum_{\ell} \sigma^x_\ell  + \cdots,
\end{equation}
with $h, U>0$, {with $U \sim 1$.  The ``$\cdots$'' are further terms consistent with the symmetries and gauge invariance, that for convenience of discussion we take to be small.  It is important to note that the total number of fermionic partons is not fixed, although the number of partons in each plane normal to the Cartesian coordinate directions is conserved modulo two.  This follows from taking the product of $\hat{G}_c = 1$ over such a plane $P$; the $B_c$ operators cancel out and we obtain the condition $(-1)^{\sum_{c \in P} (\hat{n}_c + n^0_c ) } = 1$.  If $P$ has an even number of lattice sites, this implies that $\sum_{c \in P} \hat{n}_c  = 0 \operatorname{mod} 2$.}

Let us consider two limiting cases of this gauge theory. For {$h\ll 1$}, it is clear that we should focus on satisfying the $U$ term first by appropriately choosing matter field configurations. Evidently, the minimum of energy is achieved by taking $\hat{n}_c = 0$ on each cube, {and $U > 0$ ensures that there is a gap to excitations carrying gauge charge.  Below this gap, imposing the gauge constraint is then equivalent to demanding that $B_c = (-1)^{n^0_c} = -1$ on each cube.  In this limit we thus recover the pure generalized gauge theory obtained by projection in Sec.~\ref{sec:odd} from $H_{\text{X}, oc}$.}

{Turning now to the opposite limit  $h\gg 1$, the dominant term in the Hamiltonian is minimized by setting $\sigma^x_\ell = 1$ on every link, so that $B_c =1$ trivially for each $c$; Therefore, requiring $\hat{G}_c=1$ is equivalent to fixing $(-1)^{\hat{n}_c} = (-1)^{n^0_c} = -1$. This implies that $\hat{n}_c = 1$, so the Hilbert space in this limit reduced to that of a single spin-$1/2$ moment on each cube. This implies that this generalized gauge theory can emerge as a low-energy effective description of such a spin model, which of course has an odd number of spin-$1/2$ moments per unit cell.}

A similar analysis can also be performed for the odd gauge theories with a background of lineon excitations in the ground state.  In this case, we introduce  three species $f^{(\mu)}_{v,\alpha}$ of spin-$1/2$ fermions on each vertex of the cubic lattice, representing the lineons of the theory.  The index $\alpha = \uparrow,\downarrow$ still represents spin, and $\mu$ runs over $x$,$y$,$z$.  We impose the three gauge constraints $\hat{G}_v^\mu = 1$ on each vertex, with:
\begin{equation}
\hat{G}_v^x = (-1)^{(\hat{n}_v^y + \hat{n}_v^z)}A_v^x
\end{equation}
\begin{equation}
\hat{G}_v^y = (-1)^{(\hat{n}_v^x + \hat{n}_v^z + n_v^{0,x})}A_v^y
\end{equation}
\begin{equation}
\hat{G}_v^z = (-1)^{(\hat{n}_v^x + \hat{n}_v^y + n_v^{0,x})}A_v^z
\end{equation}
where $\hat{n}_v^\mu = \sum_\alpha f^{\dagger,(\mu)}_{v,\alpha}f^{(\mu)}_{v,\alpha}$.  Note that $\hat{G}_v^x\hat{G}_v^y = \hat{G}_v^z$, so these actually constitute only two independent constraints.  We have also chosen to introduce a static background charge {$n_v^{0,x} = 1$} of $x$-directed lineons.  (A background of $y$- or $z$-directed lineons could have been introduced along similar lines.)

{We consider the Hamiltonian
\begin{equation}
H_{\text{gauge},v} = -\sum_c B_c + U'\sum_{\mu,v,\alpha}f^{\dagger,(\mu)}_{v,\alpha}f^{(\mu)}_{v,\alpha} - h'\sum_\ell \sigma_\ell^z  + \cdots
\end{equation}
where $h', U'>0$ and $U' \sim 1$.
Once again, it is straightforward to analyze various limits of the gauge theory Hamiltonian.  For $h' \ll 1$, the $U'$-term requires $\hat{n}_v^\mu = 0$ in the ground state for all $\mu$. Below the lineon gap, as a result of the background charge,  the gauge constraints reduce to $A_v^x = 1$ and $A_v^y = A_v^z = -1$ on every vertex.  We thus obtain the pure generalized gauge theory that we obtained by projection in Sec.~\ref{sec:odd}.}

{In the other limit, with $h' \gg 1$, we have $\sigma_\ell^z = 1$ on every link, so that $A_\mu^v = 1$ for all $\mu$.  The gauge constraints then become $(-1)^{(\hat{n}_v^y + \hat{n}_v^z)} = 1$ and $(-1)^{(\hat{n}_v^x + \hat{n}_v^z)} = (-1)^{(\hat{n}_v^x + \hat{n}_v^y)} = -1$ which is satisfied by $\hat{n}_v^x = 1$ and $\hat{n}_v^y = \hat{n}_v^z = 0\,\text{mod}\,2$.  As before, the system will have a single spin-$1/2$ degree of freedom on each vertex.  This demonstrates, as proof of principle, how this odd X-cube parton theory can consistently emerge as a low-energy effective theory of a physical spin system with an odd number of spin-$1/2$ moments per site.}

\section{Dual Ising Models}
\label{sec:dual}

To study the odd X-cube models in more detail, it is useful to write down their dual Ising models.  {More precisely, we consider dualities between the pure generalized X-cube gauge theories of Sec.~\ref{sec:odd} and Ising-like models, building on earlier work of Ref.~\onlinecite{fracton2}. These dualities are analogous to that between two-dimensional $Z_2$ gauge theory and the conventional transverse field Ising model.}  In the case of an odd $Z_2$ gauge theory, the corresponding Ising dual is fully frustrated, as reviewed in Appendix~\ref{app:duality}.  For X-cube models, the dualities come in two different varieties, {depending on whether we treat the $A^v_{\mu}$ term or the $B_c$ term as the Gauss' law constraint of a generalized gauge theory.  This corresponds to working in lineon-free and fracton-free sectors, respectively.}

\subsection{Plaquette Ising Models}

One way to dualize X-cube models is to work with the generalized gauge theory obtained by projecting to the lineon-free sector.  Before proceeding to the odd X-cube model, we first review this duality in the case of the ordinary (even) X-cube model.\cite{fracton2} We begin with the even X-cube theory,  augmented by a Zeeman field term that generates hopping of planon composites of fractons:
\begin{equation}
H_{\text{X},e}'=   - \sum_c B_c  - g\sum_{\ell} \sigma^z_\ell \text{.}  \label{eq:Xyzprime}
\end{equation}
This Hamiltonian is supplemented by the constraint $A^v_{\mu}  = 1$.  Note that we have not included a $\sigma^x$ perturbation, which does not respect the constraint.  Within this sector, we can dualize the theory by solving the $A^v_\mu=1$ constraint in terms of a new set of spins-$1/2$  $\tau$ spins on the dual of the original cubic lattice. Unlike the case of the toric code (see Appendix~\ref{app:duality}), there is no simple two-spin expression for $\sigma$ which can simultaneously solve all three of these constraints.  Rather, the constraints can be solved by the following four-spin expressions:
\begin{equation}
\sigma^z_{\ell} = \prod_{i\in\square_{d,\ell }}\tau^z_i
\end{equation}
where the product is over spins $\tau_i$ at the four corners of the plaquette of the dual lattice normal to link $\ell$ on the direct lattice.  By checking commutation relations, we can also easily identify:
\begin{equation}
B_{c_i}= \prod_{\ell \in \partial c_i} \sigma^x_\ell \equiv \tau^x_i
\end{equation}
where   $i$ is the site on the dual lattice that lies at the centre of cube  $c_i$ on the direct lattice.
Using these expressions, we can then rewrite the X-cube Hamiltonian in dual form as:
\begin{equation}
\tilde{H}_{\text{X},e}'= -\sum_i \tau^x_i - g\sum_{p } \prod_{i\in p} \tau^z_i
\label{plaqdual}
\end{equation}
where the sums run over the vertices $i$ and plaquettes $p$ of the dual lattice.  This dual Hamiltonian takes the form of a plaquette Ising model in a transverse magnetic field.  The $g\rightarrow 0$ limit, which corresponds to the deconfined phase of the X-cube model, has all spins polarized in the $x$ direction.  The $g\rightarrow\infty$ limit, corresponding to the confined phase of the X-cube model, leads to an ordered phase of the Ising spins in which all plaquette terms are minimized. While the precise nature of the ordering is subtle and to our knowledge not completely characterized, it seems likely that it will involve the spontaneous breaking of the subsystem symmetries that are characteristic of plaquette Ising models.

We now wish to perform the same duality transformation for an odd X-cube gauge theory.  For concreteness, let us consider the odd theory with a uniform background density of $x$-directed lineons.  {The Hamiltonian is unchanged but the constraint is modified to $A^v_x = -A^v_y = -A^v_z = 1$.}
The constraint can once again be solved by a four-spin expression, after introducing a suitable sign structure, as follows:
\begin{equation}
\sigma^z_{\ell} = \eta_{\ell} \prod_{i\in\square_{d,\ell}}\tau^z_i
\end{equation}
where the product is over the four sites on the dual lattice plaquette  pierced by the link $\ell$ on the direct lattice. $\eta_{\ell}$ is a fixed ($i.e.$ non-dynamical) function defined on the links of the direct lattice (plaquettes of the dual lattice), taking values $1$ or $-1$, which is forced to obey the constraints:
\begin{align}
\prod_{\ell\in+_{xy}}\eta_{\ell} &= \prod_{\ell\in+_{xz}}\eta_\ell = -1,
&\prod_{\ell\in+_{yz}}\eta_\ell = 1
\label{oxcconstraint}
\end{align}
on each vertex. In words, there must be an odd number of negative $\eta$ values in both the $xy$ and $xz$ planes, with an even number of negative values in the $yz$ plane.  This can be achieved, for example, {by having $\eta_\ell = -1$}  on one $x$-directed link touching each vertex.  Putting everything together, the dual Ising Hamiltonian of the odd X-cube model is given by:
\begin{equation}
\tilde{H}_{\text{X},yz}'  = -\sum_i \tau^x_i - g\sum_{p}\eta_{p} \prod_{i\in p}  \tau^z_i
\end{equation}
where we label dual lattice plaquettes by $p$, and the product in the second term is over the sites at the corners of the plaquettes (note that we have now labeled $\eta$ by its dual-lattice plaquette index rather than its direct-lattice link index). The constraints of Eq.~\eqref{oxcconstraint} dictate that, adjacent to each site of the dual lattice, there must be either one $x$-oriented plaquette of flipped sign, or two plaquettes of flipped sign with normals $y$ and $z$.  This dual Hamiltonian once again corresponds to a plaquette Ising model in a transverse magnetic field.  However, the model is now frustrated, since all the constraints cannot be simultaneously satisfied.  We will consider the various phases of this model further in the next section.
 
 {We note that if we start with the other class of odd X-cube model, with the sign of the $B_c$ term flipped but with the coefficient of $A^v_\mu$ negative, and project to the $A^v_\mu = 1$ lineon-free subspace, we again obtain the even X-cube generalized pure gauge theory discussed above.  The only difference is the sign of the $B_c$ term, which results in an unimportant flipped sign of transverse field in the dual model.  However, the same odd X-cube model has an interesting dual if projected to the subspace $B_c = -1$, which has dynamical lineons and a frozen background of fractons.}

\subsection{Multi-Spin Ising Models}
As an alternative pathway to an Ising-like dual of the X-cube model, {we can consider the generalized gauge theory describing the fracton-free sector. Beginning with the even X-cube model, we impose the constraint $B_c = 1$ and consider the Hamiltonian takes the form:
\begin{equation}
H_{\text{X},e}'  = -\sum_{v,\mu} A^v_{\mu} - g'\sum_\ell \sigma^x_\ell  \label{eq:Xicprime}
\end{equation}
where, notably, the simplest perturbation respecting the constraint is now a $\sigma^x$ term, not $\sigma^z$.}  constraint can be solved by introducing three sets of spins, $(\tau,\mu,s)$, on each vertex of the lattice.  In terms of these new variables, the original spins can be written as:
\begin{subequations}
\begin{align}
\sigma^x_{i,i+x} &= \tau^z_i\tau^z_{i+x}\mu^z_i\mu^z_{i+x}\\
\sigma^x_{i,i+y} &= \mu^z_i\mu^z_{i+y} s^z_i s^z_{i+y}\\
\sigma^x_{i,i+z} &= s^z_i s^z_{i+z}\tau^z_i\tau^z_{i+z}
\end{align}
\end{subequations}
By checking commutation relations, we can also identify:
\begin{subequations}
\begin{align}
A^v_{x} &= \prod_{+_{yz},,v}\sigma^z = \tau^x_i \mu^x_i\\
A^v_{y} &= \prod_{+_{zx},v}\sigma^z = s^x_i \mu^x_i\\
A^v_{z} &= \prod_{+_{xy},v}\sigma^z = \tau^x_i s^x_i
\end{align}
\end{subequations}
Putting all of the pieces together, we can rewrite the Hamiltonian as:
\begin{align}
\tilde{H}_{\text{X},e}' & = -\sum_i (\tau^x_i s^x_i + \tau^x_i \mu^x_i + s^x_i \mu^x_i) \nonumber\\
&\,\,\,\,\,\,- g'\sum_i (\tau^z_i\tau^z_{i+x}\mu^z_i\mu^z_{i+x} + \mu^z_i\mu^z_{i+y} s^z_i s^z_{i+y} \nonumber\\&\,\,\,\,\,\,\,\,\,\,\,\,\,\,\,\,\,\,\,\,\,\, + s^z_i s^z_{i+z}\tau^z_i\tau^z_{i+z})
\end{align}
which is an unusual type of multi-spin Ising model.  Note that, much like the plaquette Ising model, this model has planar subsystem symmetries, such as flipping all $\tau$ spins ($\tau^z\rightarrow -\tau^z$) along any $xz$-plane.

We can now extend this duality to an odd X-cube model with a background density of fractons, {keeping the Hamiltonian the same but modifying the constraint to $B_c = -1$.}  We can solve this constraint through a similar expression to the even theory, but with an added sign structure:
\begin{subequations}
\begin{align}
\sigma^x_{i,i+x} &= \eta_{i,i+x}\tau^z_i\tau^z_{i+x}\mu^z_i\mu^z_{i+x}\\
\sigma^x_{i,i+y} &= \eta_{i,i+y}\mu^z_i\mu^z_{i+y} s^z_i s^z_{i+y}\\
\sigma^x_{i,i+z} &= \eta_{i,i+z}s^z_i s^z_{i+z}\tau^z_i\tau^z_{i+z}
\end{align}
\end{subequations}
where $\eta$ is a fixed function defined on the links of the lattice, taking values $-1$ or $1$, which satisfies the condition:
\begin{equation}
\prod_{\ell\in\partial c}\eta_\ell = -1
\end{equation}
on each cube of the lattice.  In other words, there must be an odd number of negative $\eta$ values on the links around the boundary of each cube.  In the language of the dual lattice, we can regard this as an odd number of $\eta$ values on the plaquettes touching each vertex.  Using this sign structure, the dual Ising Hamiltonian of this odd X-cube model can be written as:
\begin{align}
H = -\sum_i (&\tau^x_i s^x_i + \tau^x_i \mu^x_i + s^x_i \mu^x_i) \nonumber\\
- g'\sum_i (&\eta_{i,i+x}\tau^z_i\tau^z_{i+x}\mu^z_i\mu^z_{i+x} + \eta_{i,i+y}\mu^z_i\mu^z_{i+y} s^z_i s^z_{i+y} \nonumber\\
&+ \eta_{i,i+z}s^z_i s^z_{i+z}\tau^z_i\tau^z_{i+z})
\end{align}
which is a frustrated version of the multi-spin Ising model.

\section{Proximate Ordered Phases}
\label{sec:order}

In the previous sections, we have argued that the odd X-cube models give rise to ordered phases upon condensation of their emergent quasiparticles, as required by the Lieb-Schultz-Mattis theorem.  However, we have so far said nothing about what specific types of ordered phases are obtained.  The identification of these ordered phases represents an important task, since systems hosting such orders may be proximate to fracton phases.  In this way, mapping out the adjacent ordered phases can provide us with insight into which physical systems may give rise to fractons.  We therefore set out to identify which symmetries are broken by various condensation transitions out of the odd X-cube models.

\begin{figure}[t!]
 \centering
 \includegraphics[scale=0.25]{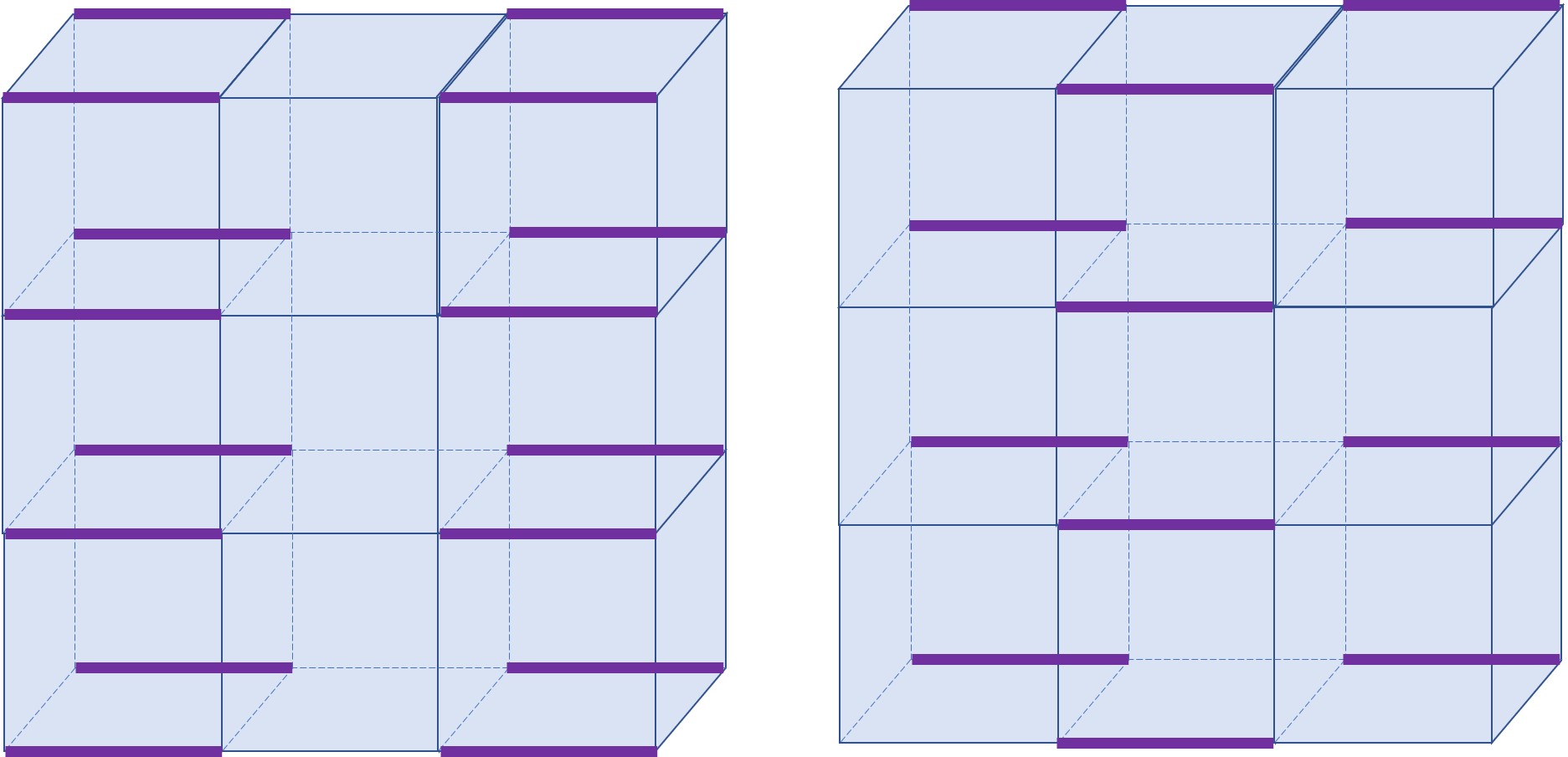}
 \caption{Two examples of bond-ordered phases proximate to the odd X-cube model with a uniform background of $x$-directed lineons.  There must be one flipped $x$-directed bond touching each vertex of the lattice.  These bond orders can come in several different columnar and staggered varieties.}
 \label{fig:bonds}
\end{figure}

We first consider an odd X-cube model with a uniform background density of lineons, which we take to be directed in the $x$-direction for concreteness.  This theory is obtained when the lineons carry the fractionalized global $U(1)$ charge, so condensation of lineons leads to ordinary superfluid (or magnetic) order.  The more interesting possibility is that we have condensation of fractons and {their composites}, which carry only fractional {crystal} momentum.  This transition can be studied via a Hamiltonian of the form (\ref{eq:Xyzprime}), reproduced here for convenience:
{\begin{equation}
H_{\text{X},e}'=   - \sum_c B_c  - g\sum_{\ell} \sigma^z_\ell \text{.}
\tag{\ref{eq:Xyzprime}}
\end{equation}
This Hamiltonian is supplemented with the constraint $A^v_z = -A^v_y = - A^v_z = 1$.

When $g$ is small, the fractons and {their composites} are gapped, and we remain in the X-cube phase.  As $g$ is increased, however, {the fractons (and their planon composites)} eventually condense and drive the system into an ordered phase.
  As $g\rightarrow\infty$, the system prefers to have all of its spins in the $\sigma^z_\ell = 1$ state, {but this cannot be achieved due to the constraint.  Instead, the energy is minimized when the number of flipped spins is as small as possible.  The least costly way to satisfy the constraint is to have exactly one  flipped spin, with $\ell$ in the $x$-direction, touching each vertex.  Assuming that quantum fluctuations of the flipped spins do not play an important role, the system will thus}  form a type of valence bond order, with all bonds oriented in the $x$-direction.  Two orders of this type are depicted in Fig.~\ref{fig:bonds}. {We leave the question of which state is selected energetically for future work.}  {The same} considerations  apply to the odd X-cube theories with background densities of other orientations of lineons.

\begin{figure}[t!]
 \centering
 \includegraphics[scale=0.25]{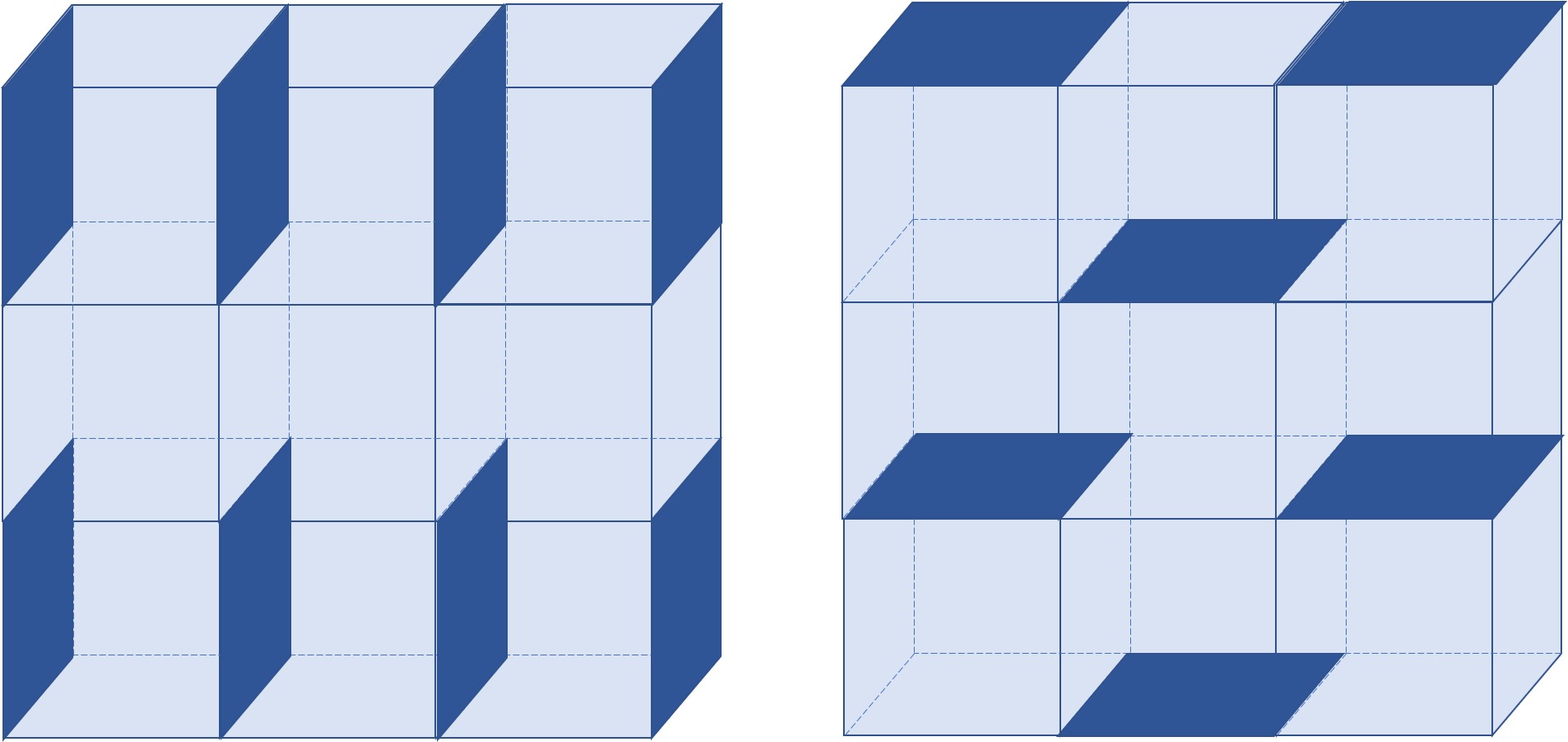}
 \caption{Two examples of plaquette-ordered phases proximate to the odd X-cube model with a uniform fracton background.  There must be one flipped plaquette touching each vertex of the lattice.  These plaquette orders can come in columnar (left), staggered (right), or other varieties.}
 \label{fig:plaqs}
\end{figure}

A more interesting case is the odd X-cube model with a uniform background density of fractons.  Such a theory tends to arise when the fractons carry the fractionalized $U(1)$ charge.  As such, condensation of fractons will lead to {a relatively mundane scenario} of a phase with $U(1)$ symmetry-breaking order.  Instead, we now consider the effect of condensing the lineons of the theory, {which can be studied via Hamiltonian (\ref{eq:Xicprime}),
\begin{equation}
H_{\text{X},e}'  = -\sum_{v,\mu} A^v_{\mu} - g'\sum_\ell \sigma^x_\ell \tag{\ref{eq:Xicprime}} \text{,}
\end{equation}
together with the constraint $B_c = -1$.}
At small $g'$, the lineons are gapped, and we remain in the X-cube phase.  As $g'\rightarrow\infty$, however, the lineons condense, and the last term in the Hamiltonian dictates most of the physics.  This term tells us that the system prefers to have as many spins as possible in the $\sigma^x_\ell = 1$ state, aligned with the transverse field.  {However, the constraint implies there} must be an odd number of flipped spins around each cube center.  This constraint can be more usefully visualized on the dual lattice, {where the spins reside on plaquettes, and it becomes the condition that}
there must be an odd number of flipped spins on the twelve plaquettes touching each \emph{vertex} of the dual lattice.  Since the $g'$ term dictates that the number of flipped spins is minimized, a ground state of the system will have precisely one flipped plaquette touching each vertex of the lattice, {which can indeed be achieved.  Models of plaquette degrees of freedom on the cubic lattice with precisely this constraint have been studied before,\cite{rsvp,sheng,cenke,zhu,moessner,zhao,albuq} and the selection of a ground state will be governed by an effective model of this kind.  One likely possibility is a plaquette solid phase; a variety of possible ordering patterns are possible, and two are depicted in Fig.~\ref{fig:plaqs}.}  How this plaquette order should be physically interpreted depends on the context.  For example, if the parent odd X-cube model arose in a fractionalized electron model, this order could be considered a type of plaquette-centered charge-density wave.

\begin{figure}[t!]
 \centering
 \includegraphics[scale=0.45]{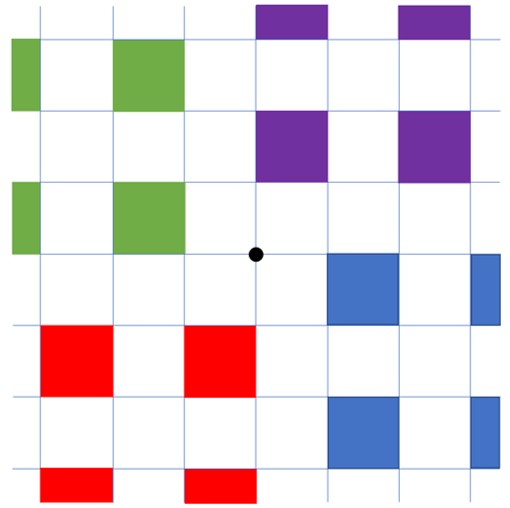}
 \caption{A vortex of a plaquette order within a two-dimensional plane, which can be regarded as a confined charge of a $U(1)$ tensor gauge theory, in close analogy with similar work in the context of valence bond order.\cite{levinsenthil}}
 \label{fig:vortex}
\end{figure}

It is of further interest to ask how a plaquette-ordered phase can be driven back into the X-cube phase.  To do this, we must consider condensing some vortex defect of the plaquette order.  Specifically, we need to condense double vortices (bound states of two identical vortices).  {To see this, it is useful to follow Ref.~\onlinecite{cenke}, where it was shown that models of plaquette degrees of freedom on the cubic lattice, with the constraint considered here, can be viewed as certain symmetric tensor gauge theories.  In Ref.~\onlinecite{yizhi}, it was shown that vortices of the plaquette order, depicted in Fig.~\ref{fig:vortex}, are charges of the gauge theory, with the following simple relationship:}
\begin{equation}
\partial_i\partial_j E_{ij} = (-1)^{x+y+z}(q_v-1)
\end{equation}
where $E_{ij}$ is a ``hollow" symmetric tensor ($i.e.$ with components $E_{xy}$, $E_{yz}$, and $E_{xz}$) which represents the plaquette variable, taking values $0$ if the plaquette is unoccupied, and $(-1)^{x+y+z}$ if it is.  The variables $x$, $y$, and $z$ denote the integer $x$-, $y$-, and $z$-coordinates of the plaquette, in units of the lattice spacing, while the variable $q_v$ represents the vortices, acting as the fracton charges of the tensor gauge theory.

In this gauge language, the most general low-energy Hamiltonian we can write for the plaquette system is:
\begin{equation}
H = \sum_p E_{ij}^2 - g\sum_{i=x,y,z}\bigg(\sum_{c}\cos(B_i)\bigg) + \sum_v q_v^2 + \ldots
\end{equation}
where $B_i = \sum_{jk} \epsilon^{ijk}\partial_jA_{ki}$ are the three gauge-invariant magnetic field operators on each cube of the lattice.  This gauge theory does not have a stable deconfined phase\cite{cenke}, which leads to the confining energy cost of vortices within the plaquette-ordered phase.  Despite the absence of a deconfined phase of this $U(1)$ tensor gauge theory, it has been shown that condensation of charge-2 objects can drive this gauge theory into the stable deconfined phase of the X-cube model.\cite{higgs1,higgs2}  We therefore conclude that condensation of doubled vortices of the plaquette order will drive the system back into the X-cube phase.

\section{Conclusions}
\label{sec:conc}

In this work, we have investigated a new type of fracton order described by {generalizations of} odd lattice gauge theories, focusing on the example of odd X-cube models.  The ground states of these theories are characterized by uniform background densities of either fractons or lineons, which leads to {phenomena analogous to crystal} momentum fractionalization in the opposite sector.  These theories are of particular interest due to the constraints of the Lieb-Schultz-Mattis theorem.  {Specifically, X-cube order can only occur at integer and half-odd-integer filling, with systems at half-odd-integer filling described by an odd X-cube gauge theory.}  This ensures that any condensation transition out of the X-cube-ordered phase will lead to some form of symmetry breaking, such as plaquette-ordered phases.  In turn, plaquette order can give rise to an X-cube phase via condensation of doubled vortices.  By identifying such proximate symmetry-breaking phases, we gain important clues as to what types of systems may host fracton order.  We also show how  odd X-cube models are related by duality transformations to various types of Ising models, such as a fully frustrated version of the plaquette Ising model.

Our work opens various further questions.  For example, the tools of this paper can be used to construct odd variants of other versions of fracton order that can occur in systems at half-odd-integer filling.  More generally, in this paper we introduced a framework for characterizing symmetry enriched fracton (SEF) phases via the action of symmetry on the fractional excitations.  We only used this framework to argue for filling constraints on X-cube fracton order, but it actually has much more general applicability, as a means to characterize SEF phases and to guide exploration of their phenomena.  We believe this will be an interesting direction for future work.  Indeed, while some attention has been devoted to symmetry-enrichment of fracton orders,\cite{yizhi18twisted,cheng,supersolid,kumar} a systematic treatment is not yet available.

\begin{acknowledgments}

We acknowledge Rahul Nandkishore for useful early discussions and Shriya Pai, Yizhi You, and Zhen Bi for insightful conversations and collaborations on related work. This work was supported by the U.S. Department of Energy, Office of Science, Basic Energy Sciences (BES) under Award number DESC0014415 (MH), and the European Union Horizon 2020 Research and Innovation Programme under European Research Council (ERC) Grant  No.~804213-TMCS (SAP).  This work was also partly supported by the Simons Collaboration on Ultra-Quantum Matter, which is a grant from the Simons Foundation (651440, MH).
\end{acknowledgments}

\begin{appendix}

\section{Failure of Flux Insertion Proof of LSM Theorems in Fracton Models\label{app:LSMfail}}

In this appendix, we explain why flux insertion arguments cannot be applied to derive LSM constraints on fracton theories, using an explicit example to illustrate where the arguments fail. To that end, consider a model of bosons at half-filling on the cubic lattice, and construct an X-cube parton effective theory along the lines of Sec.~\ref{sec:parton} where we fractionalize the boson by:  $b = a^2$.  In the X-cube phase the half-charge $a$-bosons will be in a Mott insulating state with one $a$-boson per site.  This parton theory has the gauge constraint:
\begin{equation}
B_c = (-1)^{a^\dagger a}
\end{equation}
Here we are thinking of the $a$-bosons as living on cube centers.  Now let us take a product of all the cubes $B_c$ over a $yz$ plane at fixed $x$.  Clearly this product has to be $1$ if  we are locally in the ground state everywhere in this plane.  However, the product of the right-hand side is $(-1)^{L_y L_z}$.  Note that because the $a$'s are fractons, their number in each plane is conserved modulo 2, so the product of the RHS is fixed and does not fluctuate.  Clearly, we have a contradiction if $L_y$ and $L_z$ are both odd: the only way to resolve this contradiction is to have an odd number of fractons in the $yz$-plane, i.e. an excitation above the (local) ground state.  (A similar issue arises when considering  $d=2$ $Z_2$ topological order in a similar setting, i.e. where the gauge charges are half-charge partons, and the underlying bosons are at half-filling.)

How does this affect the flux insertion arguments?  Recall that in using flux-threading argument to prove that  there must be degenerate ground states, if we turn on the background $U(1)$ vector potential along the 
$x$-direction to thread flux through the non-contractible loop in this direction, we have
\begin{equation}
\Delta P_x = 2\pi \nu L_y L_z.
\end{equation} 
For $\nu = 1/2$, we need $L_y L_z$ odd for this to be useful, which is exactly the case that lead to a contradiction  above. Therefore, there is no longer a guarantee that the system is gapped:  indeed, the fracton excitations that get forced in can combine into dipoles moving in neighboring $yz$ planes, and these can disperse, leading to a gapless spectrum.  In the absence of a  gap, we can no longer apply the flux threading argument.  In addition, a corollary of the above argument is that there is no longer a sharp separation of the Hilbert space into sectors labeled by eigenvalues of logical operators.

\section{Mathematical Details for SET and SEF Phases\label{app:math}}

Here we discuss some mathematical details that are treated lightly in Sec.~\ref{sec:lsmfracton}.  The purpose of this appendix is not to give a self-contained introduction to the relevant mathematical topics, but rather to enhance the mathematical precision of the discussion in the main text, which may be useful for some readers.

Given a group $G$, by a $G$-module we mean an Abelian group $A$ together with a homomorphism $\rho : G \to \operatorname{Aut}(A)$, by which we can view elements of $g$ as acting on elements of $A$.  We write the action of $g \in G$ on $a \in A$ as
$g a = \prescript{\rho(g)}{} a$.  To understand this notation, observe that $\rho(g) : A \to A$ is an automorphism, and we write the value of the function $\rho(g)$ as  $\prescript{\rho(g)}{} a \equiv (\rho(g))(a)$.  In contrast to the main text, here we use the symbol $A$ in place of ${\cal A}$; in the main text and later in this Appendix, ${\cal A}$ always has the physical interpretation of the group of superselection sectors (particle types), but $A$ is simply a $G$-module.  Our use of ``$G$-module'' is actually short-hand for ``left $\Z[G]$-module,'' where $\Z[G]$ is the group ring over $G$ with integer coefficients.  Elements of $\Z[G]$ are finite formal sums of elements of $G$ with integer coefficients, and the action of $G$ on $A$ given by $\rho$ makes $A$ into a left module over the ring $\Z[G]$.

Now we very briefly discuss some limited aspects of group cohomology over the $G$-module $A$.  Let $C^n(G,A)$ be the Abelian group of functions from $G^n$ to $A$, where by $G^n$ we mean the $n$-fold product of $G$ with itself.  It is important to stress that elements of $C^n(G, A)$ are merely functions, and need not be homomorphisms.  In fact, $C^n(G,A)$ is a $G$-module, coming from the $G$-module structure on $A$.  If
$\omega \in C^n (G, A)$, then we define $(g \omega) (g_1, \dots, g_n) = \prescript{\rho(g)}{} \omega(g_1, \dots, g_n)$.

There are group homomorphisms $\delta^n : C^n(G, A) \to C^{n+1}(G, A)$, that satisfy the property $\delta^{n+1} \circ \delta^n = 0$.
  We only need $\delta^1$ and $\delta^2$, which we now define:
\begin{equation}
(\delta^1 \alpha)(g_1, g_2) \equiv \alpha(g_1) + \prescript{\rho(g_1)}{} \alpha(g_2) - \alpha(g_1 g_2) \text{,}
\end{equation}
where $\alpha \in C^1(G, A)$,
and
\begin{eqnarray}
(\delta^2 \omega)(g_1, g_2, g_3) &=& \prescript{\rho(g_1)}{} \omega(g_2, g_3) + \omega(g_1, g_2 g_3) \nonumber \\
&-& \omega(g_1, g_2) - \omega(g_1 g_2, g_3) \text{.}
\end{eqnarray}
Because $\delta^2 \circ \delta^1 = 0$, we can define the Abelian group $H^2(G, A) \equiv \operatorname{ker} \delta^2 / \operatorname{im} \delta^1$.  This is one of a sequence of group cohomology groups $H^n(G, A)$.

In Sec.~\ref{sec:lsmfracton}, we emphasize the distinction between a $G$-module $A$ and its Pontryagin dual $A^*$.  Here we provide some further mathematical details.  Letting $A$ be a $G$-module, its Pontryagin dual is $A^* \equiv \operatorname{Hom}(A, {\rm U}(1))$.  That is, $A^*$ is the Abelian group of homomorphisms from $A$ to ${\rm U}(1)$.  In fact, $A^*$ also has a natural $G$-module structure that it inherits from $A$.  All we need is a left $G$-action on $A^*$, and for $\omega \in A^*$ this is given by
\begin{equation}
(g \omega)(a) \equiv \omega(g^{-1} a) \text{,}
\end{equation}
where the inverse sign is needed so that $g_1 (g_2 \omega) = (g_1 g_2) \omega$, as required for a left-action.  

Forgetting about $G$-module structure for a moment, if $A$ is a finite Abelian group, then it is a basic fact in group theory that $A \cong A^*$, \emph{i.e.} $A$ is isomorphic to its dual.  To see this one can first show that $\Z_n \cong (\Z_n)^*$, and then use the result that finite Abelian groups are products of $\Z_n$ factors.  The isomorphism between $A$ and $A^*$ is not canonical, in the sense that constructing an isomorphism requires making an arbitrary choice of generators.

Now let us consider the physical context of Abelian SET phases with symmetry group $G$, where ${\cal A}$ is the group of superselection sectors and is a finite $G$-module.  In this context, there is a canonical isomorphism  $\theta : {\cal A} \to {\cal A}^*$, given by the braiding statistics of the Abelian anyons.  $\theta$ is defined by $a \mapsto \theta_a \in {\cal A}^*$, where $\theta_a(b)$, for $b \in {\cal A}$, is the statistical phase obtained when $a$ is braided around $b$.  That is, $\theta_a(b)$ is the mutual statistical phase of the $a$ and $b$ anyons.  Physically we have $\theta_a(b + c) = \theta_a(b) \theta_a(c)$, because the phase obtained upon braiding $a$ around the fusion composite $b +c$ is the same as multiplying the phases obtained upon braiding $a$ separately around $b$ and around $c$.  Mathematically, this says that $\theta$ is a homomorphism.  Moreover, the principle of braiding non-degeneracy amounts to the statement that $\theta$ is injective. Therefore, since ${\cal A}$ and ${\cal A}^*$ are finite and of the same size, $\theta$ must be a group isomorphism.

In fact, $\theta$ is an isomorphism of $G$-modules, which follows from the physical requirement that acting with $g$ does not change the statistics of a pair of particles. Mathematically this is expressed $\theta_{g a}(g b) = \theta_a (b)$.  Putting $b \to g^{-1} b$, we have 
$\theta_{g a}(b) = \theta_a( g^{-1} b) = (g \theta_a)(b)$, which is the statement that $\theta$ is a $G$-module isomorphism.  Therefore, there is no difference between the cohomology groups $H^2(G, {\cal A})$ and $H^2(G, {\cal A}^*)$, and while one can attach different physical interpretations to these two cohomology groups, it does not really matter which one we consider.

As noted in Sec.~\ref{sec:lsmfracton}, the situation is different in gapped fracton phases, where the group of superselection sectors ${\cal A}$ is infinite.  In this case, ${\cal A}$ and ${\cal A}^*$ are not expected to be isomorphic, and it matters which of these $G$-modules we choose when using group cohomology to describe the data of an SEF phase.  In the main text we argued on physical grounds that ${\cal A}^*$ is the proper choice, and one needs the cohomology $H^2(G, {\cal A}^*)$.  Here, we first establish that when ${\cal A}$ is a countably infinite sum of $\Z_2$'s, ${\cal A}$ is countable while ${\cal A}^*$ is uncountable, so ${\cal A}$ and ${\cal A}^*$ are not isomorphic even as Abelian groups, let along as $G$-modules.  We then show that in the X-cube phase -- indeed, for any fracton phase that can be realized by a commuting Pauli Hamiltonian -- ${\cal A}$ is indeed isomorphic to a countably infinite sum of $\Z_2$'s.

Suppose that ${\cal A} = \bigoplus_{n \in \N} \Z_2$, so we have a countably infinite sum of $\Z_2$'s indexed by the natural numbers.  Here we are only interested in ${\cal A}$ as an Abelian group; we do not need to consider any $G$-module structure. Denote the generator of the $n$th summand by $a_n$, then elements $a \in {\cal A}$ are finite formal sums $a = \sum_{n \in \N} c_n a_n$, with coefficients $c_n \in \Z_2 = \{ 0, 1 \}$. To see that ${\cal A}$ is countable, given $a \in {\cal A}$ we can define $i(a) = \sum_{n \in \N} c_n n$, which is well-defined because only finitely many of the $c_n$ are non-zero.  It is easy to see that given a fixed $k \in \N$, only finitely many elements $a \in {\cal A}$ have $i(a) = k$.  Therefore we can enumerate the elements of ${\cal A}$ by first enumerating those with $i(a) = 0$, then those with $i(a) = 1$, and so on.

Now we consider ${\cal A}^*$.  An element $\varphi \in {\cal A}^*$ is completely determined by its values on generators, 
$\varphi(a_n) \in \{ +1, -1 \}$.  Any such choice of values is possible, so elements of ${\cal A}^*$ correspond bijectively to infinite sequences of $\pm 1$ entries $(\varphi(a_1), \varphi(a_2), \dots)$.  As a set ${\cal A}^*$ is therefore clearly in bijective correspondence with the set of infinite sequences of zeros and ones, indexed by the natural numbers.  Such infinite bit strings can for instance be used to uniquely represent real numbers in the interval $[0, 1]$ as binary decimals, and this makes it clear that ${\cal A}^*$ is uncountable, since $[0,1]$ is also.  Therefore it is impossible for ${\cal A}$ and ${\cal A}^*$ to be isomorphic as groups, since they are not even in bijective correspondence as sets.

To complete this discussion, it only remains to show that ${\cal A}$ for the X-cube model is isomorphic (as a group) to $\bigoplus_{n \in \N} \Z_2$.  For fracton models realized as commuting Pauli Hamiltonians (including the X-cube model), Refs.~\onlinecite{haah13commuting,haah16algebraic,fusion} obtain ${\cal A}$ as a quotient of ${\cal E}$, the group of configurations of excitations above the ground state, which is a countably infinite direct sum of $\Z_2$'s.  One takes the quotient by those excitations that can be created by operators of bounded support.  It is easy to see that the quotient ${\cal A}$ is countable (because ${\cal E}$ is), and in fracton models ${\cal A}$ is infinite.  So far we have seen that ${\cal A}$ is countably infinite, and every non-identity element in ${\cal A}$ is of order two (because this holds in ${\cal E}$).  Any such group is isomorphic to $\bigoplus_{n \in \N} \Z_2$.  To see this, let
$b_1, b_2, \dots \in {\cal A}$ be an enumeration of the non-identity elements of ${\cal A}$.   We construct from this enumeration a set $\{ a_1, a_2, \dots \}$ of independent generators of ${\cal A}$, which gives the desired isomorphism.  Start by putting $a_1 = b_1$ and $a_2 = b_2$; we know $b_2$ is not a linear combination of $b_1$ because $b_1 \neq b_2$.  Now, if $b_3$ is not a linear combination of $b_1$ and $b_2$, then put $a_3 = b_3$.  Otherwise, we choose $a_3$ to be the next $b_i$ that is not a linear combination of $a_1$ and $a_2$.  We keep going in the same manner, with \emph{e.g.} $a_4$ chosen as the next $b_i$ that is not a linear combination of $a_1, a_2, a_3$.  This clearly results in a set of independent generators for ${\cal A}$, as desired.

We note the result that ${\cal A}$ and ${\cal A}^*$ are not isomorphic groups is quite general -- it holds for any fracton phase realized as a commuting Pauli Hamiltonian.  Indeed, we expect this result holds for all gapped Abelian fracton phases.

\section{Review of Ising Duality for two-dimensional $Z_2$ gauge theory\label{app:duality}}
Here we review the Ising duality mapping~\cite{WegnerDuality} for the conventional two-dimensional $Z_2$ gauge theory, to illustrate how the even versus odd behavior of a gauge theory manifests in the dual Ising model.  The ordinary (even) gauge theory can be written in terms of a set of spins, $\sigma$, located on the links of a square lattice, with Hamiltonian given as:
\begin{equation}
H_{Z_2} = - \sum_p B_p - g\sum_\ell \sigma^z_\ell
\label{tc}
\end{equation}
where the sums are over all plaquettes $p$ and links $\ell$ of the lattice, respectively.  The spins are subject to the local constraint $A_v = 1$ for each vertex $v$. The $A_v$ and $B_p$ operators take the form:
\begin{equation}
A_v= \prod_{\ell \in +_v} \sigma^z_\ell\,\,\,\,\,\,\,\,\,\,\,\,\,\,\,\,\,\,B_p = \prod_{\ell\in\square_p} \sigma^x_\ell
\end{equation}
where the products run over the links touching the vertex $v$, and in the perimeter of the plaquette $p$, respectively.

We can dualize this theory by solving the $A_v = 1$ constraint.  This is the lattice $Z_2$ analogue of solving $\vec{\nabla}\cdot\vec{E}=0$ by letting $\vec{E}$ be the curl of an arbitrary vector field.  Similarly, the constraint is solved by writing:
\begin{equation}
\sigma^z_{\ell\equiv(i,i+y)} = \tau^z_i\tau^z_{i-x}
\end{equation}
\begin{equation}
\sigma^z_{\ell\equiv(i,i+x)} = \tau^z_i\tau^z_{i-y}
\end{equation}
where the $\tau$ spins are located at sites on the plaquette centers.  By comparing canonical commutation relations, it is also straightforward to identify:
\begin{equation}
B_p = \prod_{\square_i}\sigma^x = \tau^x_i
\end{equation}

In this dual language, we can then rewrite the Hamiltonian of Eq.~\eqref{tc} in the following form:
\begin{equation}
\tilde{H}_{Z_2, e} = -\sum_i \tau_i^x - g\sum_{\langle ij\rangle}\tau^z_i\tau^z_j
\end{equation}
which takes the form of a transverse field Ising model.  The $g\rightarrow 0$ limit, corresponding to the deconfined phase of the gauge theory, leads to a paramagnetic phase with all spins polarized in the $x$ direction.  The $g\rightarrow\infty$ limit, corresponding to the confined phase of the gauge theory, leads to the ferromagnetic phase of the Ising model.

With our duality in hand for the even gauge theory, we can now proceed to the odd case, where the Hamiltonian is unchanged, but the constraint is modified to $A_v = -1$.  A natural interpretation is that the odd gauge theory model has a background $e$ particle on every vertex of the lattice, which leads to phase factors associated with motion of $m$ particles.  We proceed as before by solving the constraint, which is accomplished by writing:
\begin{equation}
\sigma^z_{\ell\equiv(i,i+y)} = \eta_{i,i+y}\tau^z_i\tau^z_{i-x}
\end{equation}
\begin{equation}
\sigma^z_{\ell\equiv{i,i+x}} = \eta_{i,i+x}\tau^z_i\tau^z_{i-y}
\end{equation}
where $\eta$ is a non-dynamical function defined on the links, taking values $1$ or $-1$, which for each vertex $v$ must satisfy
\begin{equation}
\prod_{\ell \in +_v} \eta_\ell = -1  \text{.}  \label{eqn:eta}
\end{equation}

The dual form of the Hamiltonian Eq.~\eqref{tc} is then
\begin{equation}
\tilde{H}_{Z_2,o} = -\sum_i \tau_i^x - g\sum_{\langle ij\rangle}\eta_{ij}\tau^z_i\tau^z_j \text{,}
\label{ffim}
\end{equation}
where, due to Eq.~(\ref{eqn:eta}), $\eta$ must have a flipped sign on an \emph{odd} number of links in each plaquette of the dual lattice.
This leads to frustration in the spin system, since it is not possible to simultaneously have each bond in its energetically preferred configuration.  The Hamiltonian of Eq.~\eqref{ffim} is known as the fully frustrated Ising model~\cite{PhysRevB.30.1362}, and its link to the odd gauge theory was made in Ref.~\onlinecite{MoessnerSondhiFradkin}.

\end{appendix}

\bibliography{fracton_bib}

\end{document}